\documentclass[twocolumn,aps,prb,superscriptaddress,floatfix]{revtex4-2}

\usepackage{amsmath,amssymb}
\usepackage{graphicx}
\usepackage{hyperref}
\usepackage{xcolor}
\usepackage{listings}
\usepackage{bm}
\usepackage{booktabs}

\newcommand{\vm}{\bm{m}}

\newcommand{\mumax}{\texttt{mumax$^+$}}

\begin{document}

\title{Multimode cavity magnonics in \mumax{}: from coherent to dissipative coupling in ferromagnets and antiferromagnets}

\author{Gyuyoung Park}
\email{proslaw@kist.re.kr}
\affiliation{Center for Semiconductor Technology, Korea Institute of Science and Technology (KIST), 5 Hwarangno 14-gil, Seongbuk-gu, Seoul 02792, Korea}

\author{OukJae Lee}
\affiliation{Center for Semiconductor Technology, Korea Institute of Science and Technology (KIST), 5 Hwarangno 14-gil, Seongbuk-gu, Seoul 02792, Korea}

\author{Biswanath Bhoi}
\email{biswanath.phy@iitbhu.ac.in}
\affiliation{Nano-Magnetism and Quantum Technology Lab, Department of Physics, Indian Institute of Technology (Banaras Hindu University), Varanasi, Uttar Pradesh - 221005, India}

\date{March 4, 2026}

\begin{abstract}
Coherent coupling between microwave cavity photons and magnon excitations enables quantum transduction, magnon-mediated entanglement, and magnon number-resolved detection.
Micromagnetic simulation of photon--magnon coupling typically requires either modifying the core solver or implementing a full electromagnetic solver.
Here we present a two-tier cavity magnonics extension for \mumax{}, a GPU-accelerated open-source micromagnetic framework.
The first tier consists of CUDA kernels that integrate $N$ cavity-mode ODEs simultaneously with the LLG equation inside the GPU-based RK45 adaptive time-stepper, eliminating per-step GPU$\leftrightarrow$CPU transfers; spatially resolved mode profiles $u_n(\mathbf{r})$ enter both the coupling (weighted magnetization average via a CUDA reduction kernel) and the feedback ($H_\mathrm{rf}(\mathbf{r},t) = \sum_n h_{0,n}\,u_n(\mathbf{r})\,\mathrm{Re}[a_n(t)]$), enabling selective addressing of non-uniform spin-wave modes once integrated into the upstream build.
The second tier is a lightweight Python co-simulation class that reproduces the same uniform-mode physics through operator-split RK4 integration without recompilation; all benchmark simulations in this work use this tier.
The co-rotating coupling convention $m^{-} = m_x - im_y$ and the self-consistent field conversion $h_0 = 2g/\gamma$ ensure physically correct Rabi dynamics in both tiers.
We validate the implementation with eight benchmark simulations:
(i)~magnon-polariton anticrossing spectra (RMSE = 17~MHz, 17\% of $2g$),
(ii)~vacuum Rabi oscillations (period error $< 6$\% across three coupling strengths),
(iii)~the cooperativity phase diagram spanning weak-to-strong coupling regimes, with resolved splitting for $C > 1$ (measured 21.3~MHz at $C = 6.7$, theory 20.0~MHz),
(iv)~cavity mode-profile-dependent coupling selection rules,
(v)~multi-mode polariton hybridization with magnon-mediated cavity--cavity energy transfer,
(vi)~mode-selective coupling via spatial overlap engineering,
(vii)~antiferromagnetic magnon--cavity coupling with N\'{e}el-vector spectroscopy, and
(viii)~abnormal anticrossing from dissipative photon--magnon coupling, demonstrating the transition from level repulsion to level attraction.
Numerical convergence is verified through an operator-splitting time-step study ($\Delta t = 0.25$--$4$~ps, $< 3$\% error for all step sizes).
\end{abstract}

\maketitle

\section{Introduction}
\label{sec:intro}

\begin{figure*}[t]
  \centering
  \includegraphics[width=\textwidth]{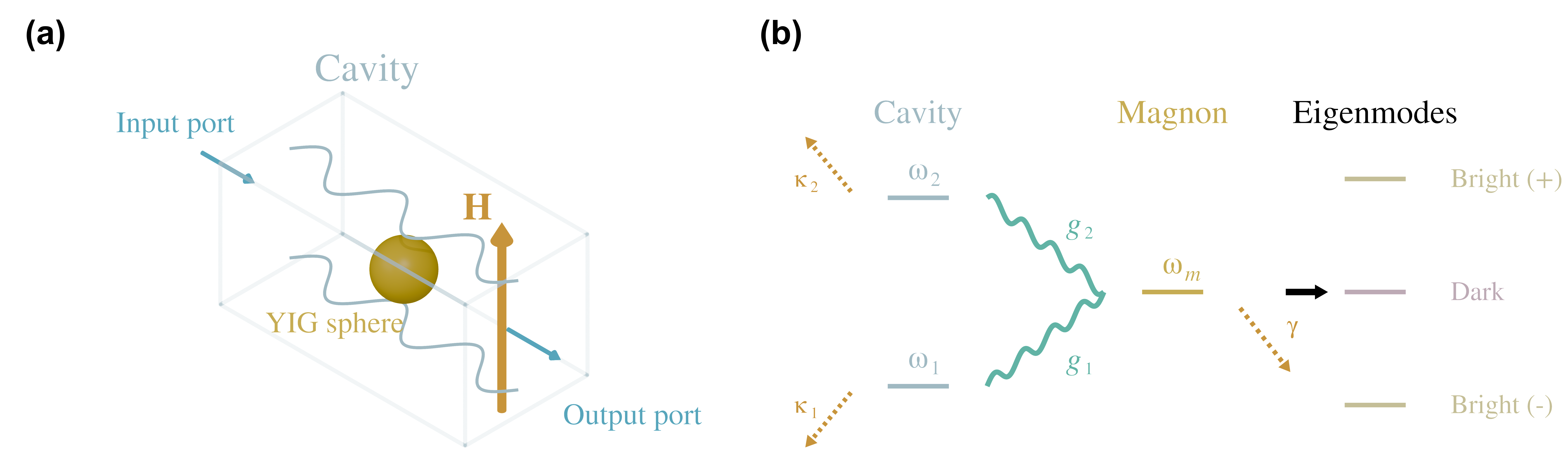}
  \caption{(a)~3D microwave cavity with a ferrimagnetic YIG sphere at the center.  A standing-wave RF field (sinusoidal curves) couples to the uniform magnon precession.  A static external field $\mathbf{H}$ along $\hat{z}$ sets the magnon frequency $\omega_m = \gamma B_0$ ($B_0 = \mu_0 H$).  Microwave signals enter and exit through coupling ports.  (b)~Energy-level diagram for the two-mode cavity--magnon system.  Two cavity modes ($\omega_1$, $\omega_2$) couple to a single magnon mode ($\omega_m$) with strengths $g_1$, $g_2$; dashed arrows denote dissipation ($\kappa_1$, $\kappa_2$, $\gamma$).  The hybridized eigenmodes consist of two bright polariton branches and one dark mode $|D\rangle \propto g_2|1\rangle - g_1|2\rangle$ that decouples from the magnon.}
  \label{fig:schematic}
\end{figure*}

Microwave photons in high-$Q$ cavities hybridize coherently with magnon excitations in ferrimagnetic materials, forming magnon-polariton quasiparticles whose spectral splitting, temporal dynamics, and spatial mode structure are all tunable via an external magnetic field~\cite{Soykal2010,Huebl2013,Tabuchi2014,Zhang2014,LachanceQuirion2019,ZareRameshti2022}.
Yttrium iron garnet (YIG) spheres [Fig.~\ref{fig:schematic}(a)] are the standard platform: the low Gilbert damping ($\alpha \sim 10^{-4}$) yields cooperativities $C = g^2/(\kappa\gamma_m) \gg 1$, placing the system firmly in the strong coupling regime~\cite{Tabuchi2014,Zhang2014,Goryachev2014}.
Experimental applications now range from coherent magnon--qubit coupling~\cite{Tabuchi2015} and single-magnon detection~\cite{LachanceQuirion2020} to dissipative magnon--photon interactions~\cite{Harder2018b} and exceptional-point physics~\cite{Zhang2017}.
Richer physics emerges when multiple cavity modes participate: dark polariton states---superpositions of cavity modes decoupled from the magnon---mediate indirect photon--photon energy transfer through virtual magnon exchange~\cite{Zhang2015,Lambert2015}, and the spatial profile of each mode selects which spin-wave modes are addressed through an overlap integral~\cite{Cao2015,Bourhill2016,Flower2019}.
Simulating these phenomena requires integrating cavity photon dynamics self-consistently with micromagnetic magnetization dynamics, which is the focus of the present work.

From a simulation perspective, capturing photon--magnon coupling requires integrating the cavity photon dynamics alongside the magnetization dynamics.
Two computational approaches exist in the literature: (i)~embedding a lumped cavity ordinary differential equation (ODE) into the time-stepping loop of a micromagnetic solver~\cite{MartinezLosa2024}, and (ii)~coupling a full finite-difference time-domain (FDTD) electromagnetic solver to the LLG equation~\cite{Yao2022}.
The first approach is physically transparent and computationally lightweight; the second captures electromagnetic retardation and spatial mode structure but is orders of magnitude more expensive.
Recently, Mart\'{i}nez-Losa del Rinc\'{o}n \textit{et al.}~\cite{MartinezLosa2024} extended \texttt{mumax$^3$} with a cavity coupling module, validating the ODE-based approach in a GPU-accelerated micromagnetic context.

In this work, we present a two-tier cavity magnonics extension for the \mumax{} framework~\cite{Moreels2026}, a recently published extensible GPU-accelerated micromagnetic framework building on the \texttt{mumax$^3$} legacy~\cite{Vansteenkiste2014}.
The first tier consists of custom CUDA kernels that integrate the cavity ODE(s) alongside the LLG equation inside the GPU adaptive time-stepper, providing native performance without per-step host transfers.
We implement a multi-mode cavity solver (\texttt{multimode\_cavity.cu}) supporting $N$ independently coupled modes and a spatially resolved mode-profile module (\texttt{cavity\_mode\_profile.cu}) that enables selective coupling to non-uniform spin-wave modes through CUDA reduction kernels.
The second tier is a lightweight Python co-simulation class that exploits the \mumax{} Python API---specifically the spatially averaged magnetization readout and a writable applied-field handle---to reproduce the same physics via operator-split RK4 integration on the CPU, providing a zero-compilation prototyping path.
Both tiers share identical physics: co-rotating coupling to $m^{-}$, the self-consistency relation $h_0 = 2g/\gamma$, and support for coherent and dissipative coupling channels.
Compared with the C-level extension of Ref.~\cite{MartinezLosa2024}, the CUDA-native tier offers comparable GPU-resident performance while the Python tier trades marginal performance for rapid deployment.

After developing the theoretical framework (Sec.~\ref{sec:model}) and describing the two-tier implementation---CUDA-native solver and Python co-simulation (Sec.~\ref{sec:impl})---we validate the implementation against analytical theory through eight benchmark simulations (Sec.~\ref{sec:benchmarks}).

\section{Theoretical framework}
\label{sec:model}

\subsection{Single-mode cavity--magnon coupling}

We consider a single-mode microwave cavity of frequency $\omega_c$ coupled to the uniform (Kittel) magnon mode of frequency $\omega_m$ in a ferrimagnetic sample~\cite{Soykal2010,Harder2018}.
In the rotating-wave approximation (RWA), the cavity photon mode $a \sim e^{-i\omega t}$ co-rotates with the transverse magnetization component $m^{-} = m_x - im_y \sim e^{-i\omega t}$, rather than the counter-rotating $m^{+} = m_x + im_y \sim e^{+i\omega t}$.
The coupled dynamics in the input-output formalism read~\cite{LachanceQuirion2019,Harder2021}
\begin{align}
  \frac{da}{dt} &= (-i\omega_c - \kappa)\,a + i g\, m^{-} + \eta(t),
  \label{eq:cavity_ode} \\
  m^{-}(t) &= m_x(t) - i\, m_y(t),
  \label{eq:mminus}
\end{align}
where $a(t)$ is the complex cavity amplitude, $\kappa = \omega_c/(2Q)$ is the cavity amplitude decay rate, $g$ is the magnon--photon coupling strength, and $\eta(t)$ is an external microwave drive.
Throughout this paper, all frequencies, coupling strengths, and decay rates are expressed in angular-frequency units (rad/s).

The cavity feeds back to the magnetization through an RF magnetic field applied along $\hat{x}$:
\begin{equation}
  H_\mathrm{rf}(t) = h_0\, \mathrm{Re}[a(t)],
  \label{eq:hrf}
\end{equation}
where $h_0$ is a conversion factor (T per unit cavity amplitude) and $\gamma = 1.76 \times 10^{11}$~rad/(s\,T) is the gyromagnetic ratio.
Physically, $h_0$ converts the dimensionless cavity amplitude ($|a|^2$ proportional to the intracavity photon number) to the real RF magnetic field acting on the sample.  It is set by the zero-point magnetic-field fluctuation of the cavity mode, $h_\mathrm{zpf} = \sqrt{\mu_0\hbar\omega_c/(2 V_\mathrm{mode})}$, where $V_\mathrm{mode}$ is the effective mode volume, and the collective spin enhancement: $g = \gamma h_\mathrm{zpf}\sqrt{2sN_s}$ with $s$ the effective spin and $N_s$ the number of participating spins~\cite{Soykal2010,ZareRameshti2022}.  A smaller mode volume or larger spin ensemble thus increases both $g$ and $h_0$.
Self-consistent coupling requires the back-action field strength to match the forward coupling $g$.
To derive $h_0$, we linearize the LLG equation about the saturated ground state $\vm \approx \hat{z}$ (small-angle precession, $m_z \approx 1$) with $\vm = (m_x, m_y, 1)$ and $m^{-} = m_x - i m_y$ as the normalized transverse magnetization amplitude ($|\vm| = 1$).
The linearized equation of motion for $m^{-}$ in the presence of the static field $B_0 \hat{z}$ and the transverse RF field $H_\mathrm{rf}\hat{x}$ reads~\cite{ZareRameshti2015,Harder2021}
\begin{equation}
  \dot{m}^{-} \approx -i\omega_m\, m^{-} + \frac{\gamma h_0}{2}\, a,
  \label{eq:mdot}
\end{equation}
where the factor of $1/2$ arises from decomposing $\mathrm{Re}[a] = (a + a^*)/2$ and retaining only the co-rotating component $a \sim e^{-i\omega t}$ within the rotating-wave approximation.
The cavity-to-magnon coupling rate is therefore $\gamma h_0 / 2$.
Note that the magnon decay rate $\gamma_m = \alpha\omega_m$ does not appear explicitly in the linearized Eq.~\eqref{eq:mdot}; in the co-simulation it is instead handled self-consistently by the full nonlinear LLG equation integrated on the GPU, which contains the Gilbert damping torque $\alpha\,\vm \times \dot{\vm}$.  The magnon dissipation thus enters the coupled dynamics through the micromagnetic solver rather than as an additional phenomenological term in the magnon equation of motion.
Equating this to the forward (magnon-to-cavity) coupling $g$ in Eq.~\eqref{eq:cavity_ode} yields the self-consistency condition
\begin{equation}
  h_0 = \frac{2g}{\gamma}.
  \label{eq:h0}
\end{equation}
This condition assumes reciprocal coupling, i.e., equal magnon-to-cavity and cavity-to-magnon interaction strengths.  In systems with nonreciprocal photon--magnon coupling---for example, those mediated by dissipative or interfacial Dzyaloshinskii--Moriya interactions~\cite{Harder2018b}---separate forward and backward coupling coefficients $g_\mathrm{f} \neq g_\mathrm{b}$ must be introduced, with $h_0 = 2g_\mathrm{b}/\gamma$ ensuring self-consistency for the backward (cavity-to-magnon) channel.

\subsection{Polariton eigenfrequencies}

Linearizing Eqs.~\eqref{eq:cavity_ode}--\eqref{eq:hrf} about the magnetic ground state yields two hybrid magnon-polariton modes with eigenfrequencies~\cite{Harder2018,ZareRameshti2022}
\begin{equation}
  \omega_\pm = \frac{\omega_c + \omega_m}{2}
    \pm \sqrt{g^2 + \left(\frac{\omega_c - \omega_m}{2}\right)^{\!2}}.
  \label{eq:polariton}
\end{equation}
At the degeneracy point $\omega_c = \omega_m$, the minimum frequency splitting is $2g$, the vacuum Rabi splitting~\cite{Soykal2010}.
Equation~\eqref{eq:polariton} is the Hermitian (lossless) limit.  Including dissipation yields complex eigenvalues obtained by the replacements $\omega_c \to \omega_c - i\kappa$ and $\omega_m \to \omega_m - i\gamma_m$:
\begin{equation}
  \tilde{\omega}_\pm = \frac{\omega_c\!-\!i\kappa + \omega_m\!-\!i\gamma_m}{2}
    \pm \sqrt{g^2 + \!\left(\frac{\omega_c\!-\!i\kappa - \omega_m\!+\!i\gamma_m}{2}\right)^{\!2}}.
  \label{eq:polariton_complex}
\end{equation}
The real parts $\mathrm{Re}(\tilde{\omega}_\pm)$ give the polariton frequencies and exhibit an avoided crossing in the strong-coupling regime, while the imaginary parts $\mathrm{Im}(\tilde{\omega}_\pm)$ describe the hybridized linewidths: far from resonance one mode is cavity-like (linewidth $\sim\kappa$) and the other magnon-like ($\sim\gamma_m$), and they exchange character at the crossing point.  For $g \gg |\kappa - \gamma_m|$ the lossless expression Eq.~\eqref{eq:polariton} approximates the real parts to order $(\kappa + \gamma_m)^2/g^2$.

\subsection{Cooperativity and coupling regimes}

The cooperativity~\cite{LachanceQuirion2019,Goryachev2014}
\begin{equation}
  C = \frac{g^2}{\kappa \gamma_m},
  \label{eq:cooperativity}
\end{equation}
where $\gamma_m = \alpha\omega_m$ is the magnon amplitude decay rate, determines the coupling regime.
For $C < 1$ (weak coupling), the mode hybridization is masked by dissipation and no spectral splitting is resolved.
For $C > 1$ (strong coupling), the splitting $2g$ exceeds the combined linewidths and a resolved anticrossing appears~\cite{Tabuchi2014,Zhang2014}.
Cooperativities exceeding $10^7$ have been achieved in high-$Q$ YIG-sphere systems~\cite{Bourhill2016}.

\subsection{Kittel frequency}

For a uniformly magnetized sphere in a static external field $B_0$ along the magnetization axis, the uniform precession (Kittel) mode frequency is~\cite{Kittel1948}
\begin{equation}
  \omega_m = \gamma B_0.
  \label{eq:kittel}
\end{equation}
Sweeping $B_0$ tunes $\omega_m$ through $\omega_c$, producing the characteristic anticrossing in the frequency spectrum.
Higher-order magnetostatic modes (Walker modes~\cite{Walker1957}) can also couple to the cavity but are not considered in the present uniform-mode simulations.

\subsection{Spatial mode profiles}
\label{sec:mode_profiles}

When the magnetic sample spans a significant fraction of the cavity mode wavelength, the RF field acquires a spatial dependence.
The effective magnon--photon coupling is then governed by the overlap integral between the cavity mode profile and the magnon mode~\cite{Cao2015,Bourhill2016,Flower2019}:
\begin{equation}
  g_\mathrm{eff} = g_0\, \langle u(\mathbf{r}) \rangle
  = g_0 \frac{1}{V}\int u(\mathbf{r})\,dV,
  \label{eq:overlap}
\end{equation}
where $u(\mathbf{r})$ is the normalized cavity mode profile and the integral is over the sample volume $V$.
For rectangular cavities, the standing-wave profiles are~\cite{ZareRameshti2015}
\begin{equation}
  u(\mathbf{r}) = \cos\!\left(\frac{n_x \pi x}{L_x}\right)
                   \cos\!\left(\frac{n_y \pi y}{L_y}\right)
                   \cos\!\left(\frac{n_z \pi z}{L_z}\right),
  \label{eq:mode_profile}
\end{equation}
where $(n_x, n_y, n_z)$ label the cavity mode.
For a sample centered in the cavity and spanning the full mode volume, the overlap integral with the spatially uniform Kittel magnon vanishes for any $n_i \neq 0$, because the positive and negative lobes cancel upon integration.
In general, the overlap depends on the sample position and size relative to the mode wavelength, and can be nonzero for higher-order modes when the sample breaks the symmetry of the cavity field~\cite{Cao2015,Bourhill2016}.
In the present simulations, the sample occupies the full grid and is centered in the mode volume, so the selection rule holds exactly.
Higher-order cavity modes can still couple to spin-wave modes of matching spatial symmetry~\cite{Cao2015,Li2019,Hou2019}.

\subsection{Multi-mode cavity extension}
\label{sec:multimode}

For $N$ cavity modes coupled to a single magnon, the cavity equations generalize to~\cite{Zhang2015,Lambert2015}
\begin{align}
  \frac{da_j}{dt} &= (-i\omega_j - \kappa_j)\, a_j + i g_j\, m^{-} + \eta_j(t),
  \label{eq:multimode_cavity} \\
  H_\mathrm{rf}(t) &= \sum_{j=1}^N h_{0,j}\, \mathrm{Re}[a_j(t)],
  \label{eq:multimode_hrf}
\end{align}
with $h_{0,j} = 2g_j/\gamma$.
For $N=2$ modes with equal coupling ($g_1 = g_2 = g$) [Fig.~\ref{fig:schematic}(b)], the $3 \times 3$ coupling matrix
\begin{equation}
  \mathcal{H} = \begin{pmatrix}
    \omega_1 & g_1 & 0 \\
    g_1 & \omega_m & g_2 \\
    0 & g_2 & \omega_2
  \end{pmatrix}
  \label{eq:3x3}
\end{equation}
yields three eigenvalues: two bright polariton branches and one dark mode $|D\rangle \propto g_2|1\rangle - g_1|2\rangle$ that decouples from the magnon at the crossing point~\cite{Zhang2015}.
The off-diagonal zeros ($\mathcal{H}_{13} = \mathcal{H}_{31} = 0$) reflect the assumption that the two cavity modes do not couple directly to each other; their interaction is mediated entirely by the shared magnon mode.  This is physically justified when the two modes belong to the same multimode cavity---distinct resonant modes of a single structure---and the YIG sphere sits within the spatial overlap region of both mode profiles, so that each mode independently couples to the magnon via its own RF field distribution [Eq.~\eqref{eq:overlap}].  If, instead, the cavity modes belong to physically separate resonators, then the magnon sample must be placed inside both cavities' field regions---e.g., at the center of a common coupling bus~\cite{Lambert2015}---and any residual direct photon--photon coupling (e.g., through evanescent fields) should be included as a nonzero $\mathcal{H}_{13}$.
The dark mode enables indirect energy transfer between cavity modes without magnon excitation, a mechanism proposed for quantum transduction~\cite{Lambert2015}.

\subsection{Antiferromagnetic magnon modes}
\label{sec:afm_theory}

In a two-sublattice antiferromagnet with exchange field $H_E$ and uniaxial anisotropy field $H_A$, the zero-field antiferromagnetic resonance (AFMR) frequency is~\cite{Keffer1952,Kampfrath2011}
\begin{equation}
  \omega_\mathrm{AFMR} = \gamma\mu_0 \sqrt{2 H_E H_A + H_A^2}.
  \label{eq:afmr}
\end{equation}
An external field $B$ along the easy axis splits this into two branches,
\begin{equation}
  \omega_{\pm} = \omega_\mathrm{AFMR} \pm \gamma B,
  \label{eq:afm_split}
\end{equation}
valid below the spin-flop transition at $B_\mathrm{sf} = \mu_0\sqrt{2 H_E H_A + H_A^2}$.
Both modes couple to the cavity through the net magnetization $\vm_\mathrm{tot} = (\vm_1 + \vm_2)/2$, giving a $3\times 3$ coupling Hamiltonian
\begin{equation}
  \mathcal{H}_\mathrm{AFM} = \begin{pmatrix}
    \omega_+ & g & 0 \\
    g & \omega_c & g \\
    0 & g & \omega_-
  \end{pmatrix},
  \label{eq:afm_3x3}
\end{equation}
whose eigenvalues yield three polariton branches.
The N\'{e}el vector $\bm{n} = (\vm_1 - \vm_2)/2$ does not couple to the uniform cavity field and thus reveals the bare magnon spectrum without hybridization gaps.
At $B = 0$, the two magnon modes are degenerate at $\omega_c$; both contribute coherently, enhancing the vacuum Rabi splitting to $2\sqrt{2}\,g$~\cite{Bialek2023}.

\section{Implementation}
\label{sec:impl}

\subsection{CUDA-native solver}
\label{sec:cuda_solver}

The first tier integrates the cavity ODE(s) directly inside the \mumax{} GPU time-stepper as a \texttt{DynamicEquation}, so that the cavity amplitude(s) and the magnetization are advanced in a single, unified RK45 step with no host-side intervention.

\paragraph{Multi-mode cavity kernel (\texttt{multimode\_cavity.cu}).}
The \texttt{MultimodeCavityRHSQuantity} class evaluates the right-hand side for $N$ cavity modes.
The state vector $(a_{1,\mathrm{re}}, a_{1,\mathrm{im}}, \ldots, a_{N,\mathrm{re}}, a_{N,\mathrm{im}}, 0)$ lives on a single-cell GPU field, and each mode's ODE [Eq.~\eqref{eq:multimode_cavity}] is evaluated per-mode on the host with the weighted magnetization average from the GPU.
The combined RF feedback field is accumulated on the GPU: modes without a spatial profile contribute a uniform field, while modes with a profile launch a per-mode CUDA kernel \texttt{k\_ScaledModeProfileField} that computes $h_{0,n}\,u_n(\mathbf{r})\,\mathrm{Re}[a_n]$ cell-by-cell.

\paragraph{Spatially resolved mode profile (\texttt{cavity\_mode\_profile.cu}).}
The mode profile $u_n(\mathbf{r})$ is stored as a scalar field on the ferromagnet grid.
A single-block CUDA reduction kernel (\texttt{k\_WeightedFieldAverage}) computes the weighted spatial average
\begin{equation}
  \langle u_n\, m_\alpha \rangle = \frac{1}{N_\mathrm{geo}} \sum_i u_n(\mathbf{r}_i)\, m_\alpha(\mathbf{r}_i),
  \label{eq:weighted_avg}
\end{equation}
replacing the uniform average in the coupling term.
A second kernel (\texttt{k\_SpatialCavityRFField}) computes the spatially varying feedback field $H_\mathrm{rf}(\mathbf{r},t) = h_0\,u(\mathbf{r})\,\mathrm{Re}[a(t)]$.
For the fundamental mode ($u = 1$), both kernels reduce to the standard uniform operations.

The CUDA kernels are complete and ready for integration into the \mumax{} build system; the required interface additions to the \texttt{Ferromagnet} class (cavity-mode accessors, \texttt{DynamicEquation} registration, and Python bindings) are specified in the header files distributed with the extension.
The benchmark simulations in this work use the Python co-simulation tier described below, which implements the same physics for the uniform-mode case; the CUDA-native tier becomes essential for large-grid simulations and for coupling to $k \neq 0$ spin-wave modes through spatially resolved profiles.

\subsection{Python co-simulation}
\label{sec:python_cosim}

The Python class \texttt{CavityMagnon} wraps the cavity ODE and a co-simulation loop, providing a compilation-free path that is used for all benchmarks in this work:

\begin{lstlisting}[caption={Python co-simulation loop (Tier~2).},label=lst:cosim]
for i in range(nsteps):
    # 1. Read magnetization from GPU
    m = magnet.magnetization.average()
    m_plus = m[0] + 1j * m[1]

    # 2. Step cavity ODE (CPU, RK4)
    #    uses m^- = conj(m^+) internally
    a = cavity.step_rk4(a, m_plus, t, dt)

    # 3. RF feedback field (h0 = 2g/gamma)
    H_rf = h0 * a.real

    # 4. Update RF field on magnet
    magnet.bias_magnetic_field = (H_rf, 0, 0)

    # 5. Advance LLG on GPU
    world.timesolver.run(dt)
\end{lstlisting}

The static field $B_\mathrm{ext}$ is set on the \texttt{world} object while the time-varying RF field is applied at the \texttt{magnet} level; since \mumax{} sums these internally, this separation avoids double-counting.
A guard clause prevents simultaneous use of both tiers, which would apply the RF field twice.
The Python tier incurs a per-step GPU$\to$CPU transfer overhead (Sec.~\ref{sec:performance}), but is sufficient for the small grids used in the present benchmarks and provides a rapid prototyping path when recompilation is impractical.

\subsection{Cavity ODE integration}

The cavity amplitude $a(t)$ is integrated with a fourth-order Runge--Kutta (RK4) scheme.
We choose RK4 over lower-order methods (e.g., forward Euler) because the cavity ODE is oscillatory at $\omega_c \sim 2\pi \times 5$~GHz; a first-order scheme would require impractically small time steps to control amplitude and phase drift over the thousands of oscillation cycles needed for spectral analysis, whereas RK4 provides fourth-order accuracy in $\Delta t$ and faithfully preserves the oscillation amplitude with the same step sizes used by the LLG solver.
During each RK4 sub-step, the magnetization $m^{-}$ is held constant (operator-splitting approximation), which is valid when $\Delta t \ll 1/\max(\omega_c, \omega_m)$.
The cavity right-hand side implements the co-rotating coupling:
\begin{lstlisting}[caption={Cavity ODE right-hand side.},label=lst:rhs]
def cavity_rhs(self, a, m_plus, t):
    m_minus = np.conj(m_plus)
    return (-1j*self.omega_c - self.kappa)*a \
           + 1j*self.g_eff * m_minus \
           + self.drive(t)
\end{lstlisting}
The conjugation in line~2 converts $m^{+} = m_x + im_y$ (read from \mumax{}) to the co-rotating $m^{-} = m_x - im_y$.

\subsection{Performance considerations}
\label{sec:performance}

The CUDA-native tier eliminates per-step host transfers entirely: the cavity amplitudes, the weighted magnetization average, and the RF feedback field all reside on the GPU and are advanced together with the LLG equation by the adaptive RK45 stepper.
The only overhead is the per-mode evaluation of the cavity ODE right-hand side, which involves one CUDA reduction (weighted average) and one field kernel (RF feedback) per mode---negligible for the typical $N \leq 8$ modes.

\begin{figure}[tb]
  \centering
  \includegraphics[width=\columnwidth]{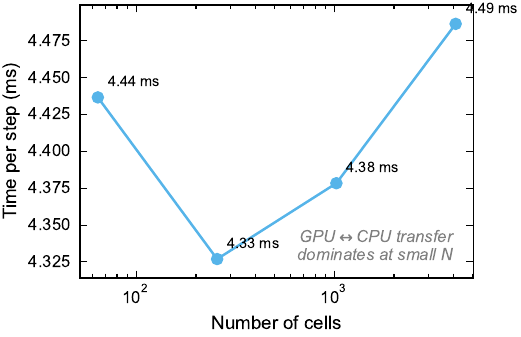}
  \caption{Computational scaling.  Wall-clock time per co-simulation step vs.\ number of grid cells (from $8\times 8$ to $64\times 64$).  The per-step cost remains approximately constant at ${\sim}4.4$~ms across the range tested, confirming that the GPU$\leftrightarrow$CPU transfer and Python overhead dominate over the GPU LLG computation for these grid sizes.}
  \label{fig:scaling}
\end{figure}

For the Python co-simulation tier, the primary overhead is the per-step GPU$\to$CPU transfer of the averaged magnetization (\texttt{magnetization.average()}).
Benchmarking the wall-clock time per co-simulation step as a function of grid size (Fig.~\ref{fig:scaling}) reveals a constant ${\sim}4.4$~ms/step from $8 \times 8$ (64 cells) to $64 \times 64$ (4096 cells), indicating that the GPU$\leftrightarrow$CPU transfer and Python overhead (${\sim}4$~ms/step) dominate over the GPU LLG computation for these small grids.
For larger grids ($256 \times 256$ and above) the GPU LLG cost would eventually exceed the host-transfer overhead, restoring GPU-limited scaling; however, such grids also couple to $k \neq 0$ spin-wave modes and therefore require the spatially resolved CUDA tier rather than the uniform Python co-simulation.
The present benchmarks use the Python tier, which is sufficient for the $8 \times 8 \times 1$ grids employed here.
The CUDA-native tier eliminates the per-step host transfer entirely, making it the appropriate path for production-scale spatially resolved simulations.

\section{Benchmark simulations}
\label{sec:benchmarks}

All simulations use YIG-like material parameters: $M_s = 140$~kA/m, $A_\mathrm{ex} = 3.5$~pJ/m, $\alpha = 3 \times 10^{-4}$~\cite{Tabuchi2014}, on an $8 \times 8 \times 1$ grid with 10~nm cell size.
Dipolar interactions are disabled so that the uniform precession frequency reduces to $\omega_m = \gamma B_0$ [Eq.~\eqref{eq:kittel}], isolating the coupling physics from shape-dependent demagnetization effects.
The magnetization is initialized along $\hat{z}$ and the adaptive time solver of \mumax{} advances the LLG at each co-simulation step.
Table~\ref{tab:params} summarizes the simulation parameters for all eight benchmarks.

\subsection{Anticrossing spectrum}
\label{sec:anticrossing}

\begin{figure*}[t]
  \centering
  \includegraphics[width=\textwidth]{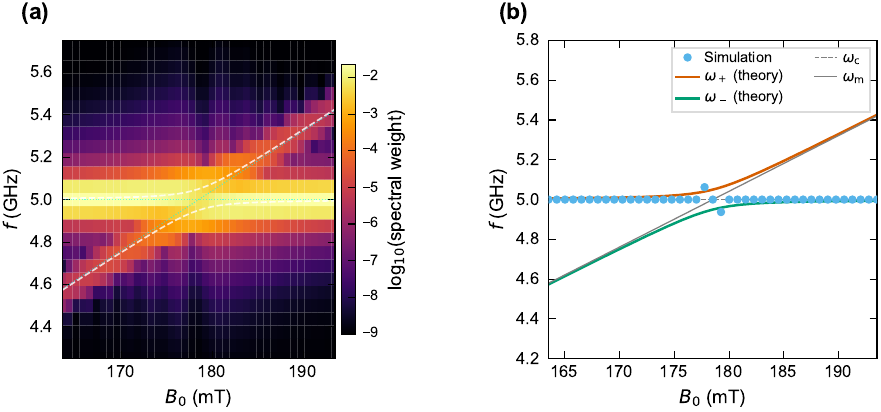}
  \caption{Magnon-polariton anticrossing.  (a)~Combined spectral weight $|$FFT$(m^{-})|^2 + |$FFT$(a)|^2$ as a function of $B_0$ and frequency.  White dashed lines: analytical polariton branches [Eq.~\eqref{eq:polariton}]; cyan dotted lines: uncoupled $\omega_c$ and $\omega_m(B_0)$.  (b)~Extracted peak frequencies (circles) compared with analytical theory (solid lines).  Parameters: $\omega_c/(2\pi) = 5$~GHz, $Q = 5000$, $g/(2\pi) = 50$~MHz.}
  \label{fig:anticrossing}
\end{figure*}

The hallmark of strong photon--magnon coupling is the avoided crossing in the frequency spectrum when the magnon frequency is swept through the cavity resonance~\cite{Huebl2013,Tabuchi2014,Zhang2014}.
We sweep $B_0$ from 164 to 194~mT in 41 steps, passing through $B_\mathrm{res} = \omega_c/\gamma \approx 179$~mT.
At each field, the cavity is excited with a short Gaussian pulse ($\sigma = 20$~ps, $t_0 = 0.1$~ns) and the coupled dynamics are evolved for 16~ns with $\Delta t = 1$~ps, giving a frequency resolution $\Delta f = 62.5$~MHz sufficient to resolve the Rabi splitting of $2g/(2\pi) = 100$~MHz.
The Gaussian temporal envelope is chosen because its Fourier transform is also Gaussian, providing a smooth broadband excitation with no spectral sidelobes; this ensures that all polariton modes within the analysis window are excited with comparable amplitude.  A sinc pulse $\eta(t) \propto \mathrm{sinc}[(t-t_0)/\tau]$ would yield a rectangular power spectrum and is in principle equally suitable, but its slowly decaying temporal oscillations can introduce truncation artifacts in finite-length simulations.
The combined spectral weight $|$FFT$(m^{-})|^2 + |$FFT$(a)|^2$ is shown in Fig.~\ref{fig:anticrossing}(a).

Two polariton branches repel each other at $B_\mathrm{res}$, with a minimum gap of $2g/(2\pi) = 100$~MHz.
The simulated spectral weight concentrates on the polariton branches and follows the analytical curves [Eq.~\eqref{eq:polariton}, white dashed lines] closely.
The extracted peak frequencies [Fig.~\ref{fig:anticrossing}(b)] track the theoretical polariton dispersion across the full field range.

\subsection{Vacuum Rabi oscillations}
\label{sec:rabi}

\begin{figure}[!htb]
  \centering
  \includegraphics[width=\columnwidth]{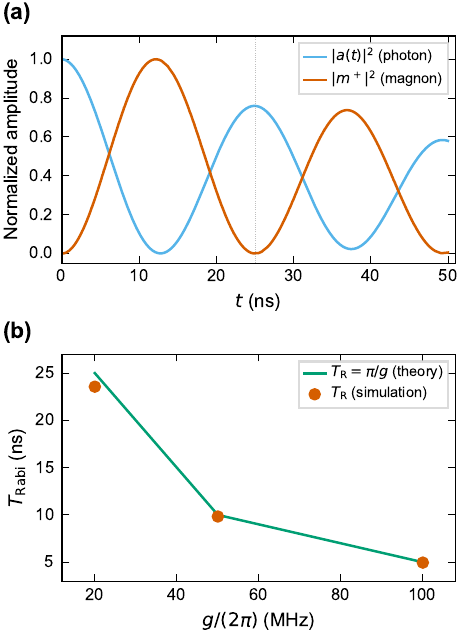}
  \caption{Vacuum Rabi oscillations.  (a)~Photon ($|a|^2$) and magnon ($|m^{-}|^2$) populations versus time for $g/(2\pi) = 20$~MHz ($T_R = 25$~ns).  Energy oscillates coherently between the cavity and magnon modes at the Rabi period $T_R = \pi/g$, with gradual decay from finite $\kappa$ and $\alpha$.  (b)~Rabi period versus coupling strength for $g/(2\pi) = 20$, 50, and 100~MHz.  Circles: simulation; line: analytical prediction $T_R = \pi/g$.}
  \label{fig:rabi}
\end{figure}

At resonance ($\omega_m = \omega_c$), the hybridized system exhibits coherent energy exchange between the photon and magnon degrees of freedom at the vacuum Rabi frequency $\Omega_R = 2g$~\cite{Soykal2010,Wolz2020}.
We initialize the cavity with $a(0) = 0.01$ and observe the time evolution.
The initial amplitude is chosen small enough ($|a(0)|^2 = 10^{-4} \ll 1$) to ensure operation deep within the linear regime where the rotating-wave approximation and the linearized coupling of Eq.~\eqref{eq:mdot} remain valid ($m_z \approx 1$, small-angle precession), while being large enough for the resulting magnetization oscillations ($\sim 10^{-4}$) to be well above the numerical noise floor of the micromagnetic solver.

Figure~\ref{fig:rabi}(a) shows the photon population $|a(t)|^2$ and magnon amplitude $|m^{-}|^2$ for $g/(2\pi) = 20$~MHz.
The two populations oscillate in antiphase with period $T_R = \pi/g = 25$~ns, as expected from the coupled-oscillator model.
The slow decay envelope reflects the finite linewidths $\kappa$ and $\gamma_m$.
Figure~\ref{fig:rabi}(b) confirms that the simulated Rabi periods agree precisely with the analytical prediction for all three tested coupling strengths ($g/(2\pi) = 20$, 50, and 100~MHz).
The agreement relies on the self-consistent coupling condition $h_0 = 2g/\gamma$ [Eq.~\eqref{eq:h0}]: without it, the Rabi frequency deviates from $2g$ because the forward (magnon$\to$cavity) and backward (cavity$\to$magnon) coupling rates are unbalanced.

\subsection{Cooperativity phase diagram}
\label{sec:cooperativity}

\begin{figure*}[t]
  \centering
  \includegraphics[width=\textwidth]{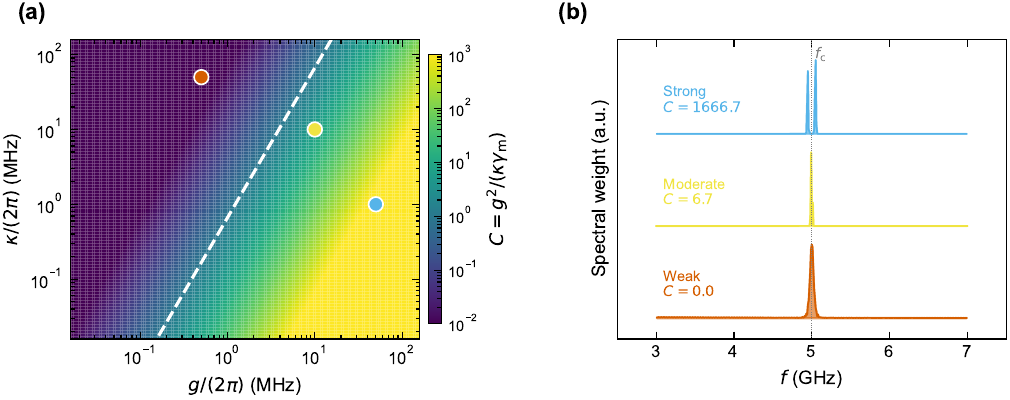}
  \caption{Cooperativity phase diagram.  (a)~Analytical cooperativity $C(g,\kappa)$ in the $(g,\kappa)$ plane; white dashed line: $C = 1$ boundary.  Circles mark the three simulation points.  (b)~Spectral weight at resonance ($B_0 = B_\mathrm{res}$) for each regime.  Strong coupling ($C = 1667$): two resolved polariton peaks separated by $2g/(2\pi) = 100$~MHz.  Moderate ($C = 6.7$): two resolved peaks with splitting $21.3$~MHz (theory: $20.0$~MHz).  Weak ($C \approx 0$): a single narrow peak with no visible splitting.}
  \label{fig:cooperativity}
\end{figure*}

The cooperativity $C$ [Eq.~\eqref{eq:cooperativity}] determines whether the photon--magnon hybridization is spectroscopically resolved~\cite{Goryachev2014,Bourhill2016}.
Figure~\ref{fig:cooperativity}(a) shows the analytically computed $C(g,\kappa)$ landscape with the $C = 1$ boundary separating the weak and strong coupling regimes.

We run simulations at three representative points [circles in Fig.~\ref{fig:cooperativity}(a)]:
\begin{itemize}
  \item \textbf{Weak coupling}: $g/(2\pi) = 0.5$~MHz, $\kappa/(2\pi) = 50$~MHz ($C \approx 0.003$).  A single narrow peak at $\omega_c$ with no splitting [Fig.~\ref{fig:cooperativity}(b), bottom].
  \item \textbf{Moderate coupling}: $g/(2\pi) = 10$~MHz, $\kappa/(2\pi) = 10$~MHz ($C = 6.7$).  Two resolved peaks at $4.991$ and $5.013$~GHz, with a measured splitting of $21.3$~MHz compared to the theoretical prediction $2g/(2\pi) = 20.0$~MHz ($6.5$\% error), confirming that $C > 1$ produces spectroscopically resolved hybridization.
  \item \textbf{Strong coupling}: $g/(2\pi) = 50$~MHz, $\kappa/(2\pi) = 1$~MHz ($C = 1667$).  Two well-separated polariton peaks spaced by $2g/(2\pi) = 100$~MHz.
\end{itemize}

The weak and strong coupling spectra are obtained by evolving for 100~ns ($\Delta f = 10$~MHz) after a broadband Gaussian pulse at resonance ($\omega_m = \omega_c$).
The moderate case requires higher frequency resolution to resolve the smaller splitting: we evolve for 200~ns and apply $4\times$ zero-padding to the FFT, giving an effective frequency grid spacing $\Delta f_\mathrm{grid} = 1.25$~MHz, well below the expected $2g/(2\pi) = 20$~MHz splitting.
The spectral signatures across all three regimes---unresolved for $C \ll 1$, resolved for $C > 1$, and strongly split for $C \gg 1$---confirm that the co-simulation correctly captures the dissipation-limited transition from weak to strong coupling~\cite{Bai2015,MaierFlaig2016}.

\subsection{Cavity mode profile effects}
\label{sec:mode_profile}

\begin{figure*}[t]
  \centering
  \includegraphics[width=\textwidth]{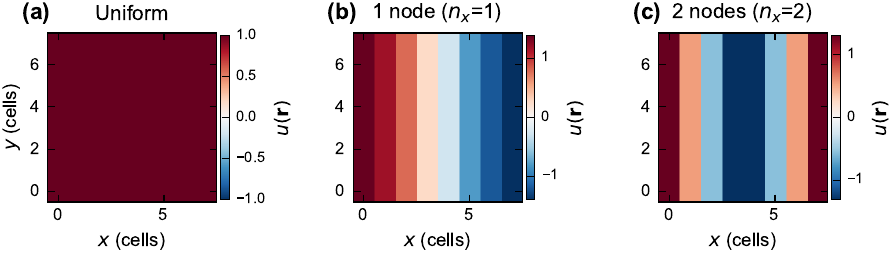}
  \caption{Spatial profiles of three cavity modes on an $8\times 8$ grid.  (a)~Uniform ($n_x = 0$): spatially constant $u(\mathbf{r}) = 1$.  (b)~One node ($n_x = 1$): $u(\mathbf{r}) = \cos(\pi x/L_x)$.  (c)~Two nodes ($n_x = 2$): $u(\mathbf{r}) = \cos(2\pi x/L_x)$.  The overlap integral $\langle u \rangle$ with the uniform Kittel magnon is 1.0 for (a) and vanishes for (b,c) due to cancellation of positive and negative lobes [Eq.~\eqref{eq:overlap}].}
  \label{fig:mode_maps}
\end{figure*}

\begin{figure*}[t]
  \centering
  \includegraphics[width=\textwidth]{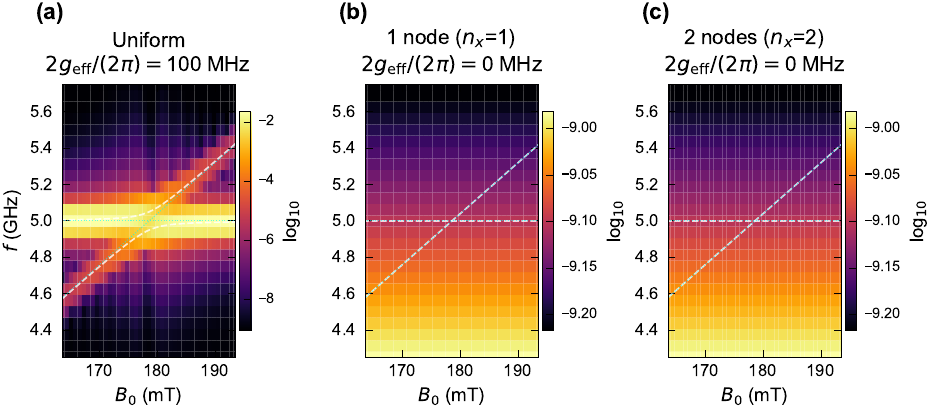}
  \caption{Mode-profile-dependent anticrossing.  Spectral weight maps for three cavity profiles coupled to the uniform Kittel magnon.  (a)~Uniform profile ($\langle u \rangle = 1$): full anticrossing with $2g_\mathrm{eff}/(2\pi) = 100$~MHz.  (b)~One-node profile ($\langle u \rangle = 0$): no anticrossing; the magnon and cavity modes cross without interaction.  (c)~Two-node profile ($\langle u \rangle = 0$): same as~(b).  Dashed lines: analytical polariton branches [Eq.~\eqref{eq:polariton}].  Parameters: $g_0/(2\pi) = 50$~MHz, $Q = 5000$.}
  \label{fig:mode_profile}
\end{figure*}

The coupling between a cavity mode and the Kittel magnon depends on their spatial overlap [Eq.~\eqref{eq:overlap}].
Figure~\ref{fig:mode_maps} shows three cavity mode profiles on the simulation grid: uniform ($n_x = 0$), one node ($n_x = 1$), and two nodes ($n_x = 2$).
The overlap integral with the uniform Kittel magnon is $\langle u \rangle = 1.00$ for the uniform mode and $\langle u \rangle = 0.00$ for both higher-order modes, consistent with the selection rule for a centered, full-volume sample~\cite{Cao2015}.

Figure~\ref{fig:mode_profile} compares the field-swept anticrossing spectra for the three mode profiles.
Only the uniform profile [panel (a)] produces an anticrossing gap ($2g_\mathrm{eff}/(2\pi) = 100$~MHz), while the higher-order profiles [panels (b,c)] show the magnon and cavity modes crossing without interaction.
As expected from the selection rule, the uniform Kittel magnon couples only to the spatially uniform cavity mode in this symmetric geometry.
Coupling to higher-order spin-wave modes with matching spatial symmetry is in principle accessible through the same framework~\cite{Li2019,Hou2019}.

\subsection{Multi-mode polariton hybridization}
\label{sec:dark_states}

\begin{figure*}[t]
  \centering
  \includegraphics[width=\textwidth]{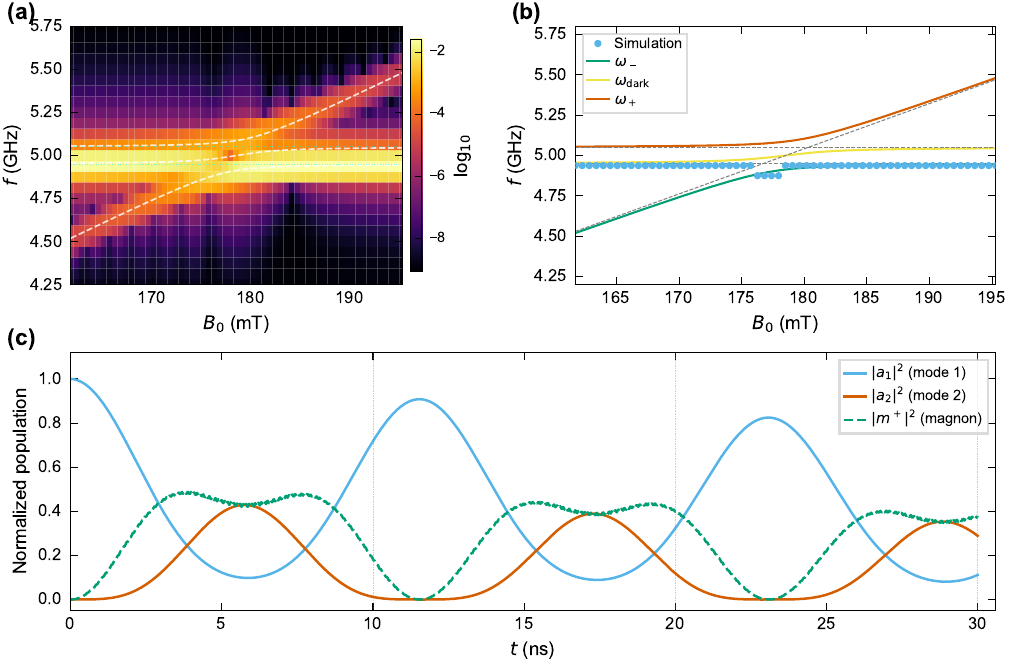}
  \caption{Multi-mode polariton hybridization with two near-degenerate cavity modes ($\omega_1/(2\pi) = 4.95$~GHz, $\omega_2/(2\pi) = 5.05$~GHz, separation $\Delta\omega/(2\pi) = 100$~MHz $= 2g$) coupled to a single magnon with $g_1/(2\pi) = g_2/(2\pi) = 50$~MHz.  (a)~Spectral weight map showing three polariton branches with overlapping avoided crossings near $B_\mathrm{res} \approx 178$~mT.  Dashed lines: eigenvalues of the $3\times 3$ coupling matrix [Eq.~\eqref{eq:3x3}].  (b)~Extracted peak frequencies (circles) versus analytical theory.  The middle branch ($\omega_\mathrm{dark}$) passes through the crossing region with minimal gap, approaching the dark-state limit as $\Delta\omega/2g \to 0$.  (c)~Multimode dynamics at the midpoint ($\omega_m = (\omega_1 + \omega_2)/2$): with only mode~1 initially excited, all three modes exchange energy periodically; the anti-phase oscillation of $|a_1|^2$ and $|a_2|^2$ demonstrates magnon-mediated cavity--cavity energy transfer.  All populations normalized to $\max|a_1|^2$.  $Q = 5000$ (anticrossing), $Q = 10\,000$ (dynamics).}
  \label{fig:dark}
\end{figure*}

When two cavity modes couple to the same magnon, the system eigenstates are hybridized polaritons whose spectrum can be obtained from the eigenvalues of Eq.~\eqref{eq:3x3}~\cite{Zhang2015,Lambert2015}.
We simulate this with two near-degenerate cavity modes at $\omega_1/(2\pi) = 4.95$~GHz and $\omega_2/(2\pi) = 5.05$~GHz (separation $\Delta\omega/(2\pi) = 100$~MHz $= 2g$), both coupled with $g/(2\pi) = 50$~MHz.
At $\Delta\omega = 2g$, the middle eigenstate contains equal contributions from both cavity modes and the magnon ($\sim$1/3 weight each).
In the strictly degenerate limit $\Delta\omega \to 0$, the magnon component vanishes and the middle eigenstate becomes the dark photonic superposition $(|a_1\rangle - |a_2\rangle)/\sqrt{2}$~\cite{Zhang2015}.

Figure~\ref{fig:dark}(a) shows the spectral weight map as $B_0$ is swept through the common resonance region near $B_\mathrm{res} \approx 178$~mT.
Three polariton branches are visible, with overlapping avoided crossings.
The dark mode appears as a branch that passes through the crossing region without hybridizing with the magnon [Fig.~\ref{fig:dark}(b)], in good agreement with the analytical eigenvalues.

At the midpoint $\omega_m = (\omega_1 + \omega_2)/2$, the multimode dynamics become visible [Fig.~\ref{fig:dark}(c)].
With mode~1 initially populated [$a_1(0) = 0.01$, $a_2(0) = 0$], all three modes exchange energy periodically: mode~1 decays, exciting the magnon, which in turn populates mode~2.
The anti-phase oscillation of $|a_1|^2$ and $|a_2|^2$ demonstrates magnon-mediated cavity--cavity energy transfer, the mechanism behind quantum transduction proposals~\cite{Zhang2015,Lambert2015,Tabuchi2015}.
At $\Delta\omega = 2g$ the magnon participates as an active intermediary with comparable amplitude; in the degenerate limit $\Delta\omega \ll 2g$, the transfer becomes effectively virtual and the magnon remains unexcited~\cite{Lambert2015}.

\subsection{Mode-selective coupling}
\label{sec:selective}

\begin{figure*}[t]
  \centering
  \includegraphics[width=\textwidth]{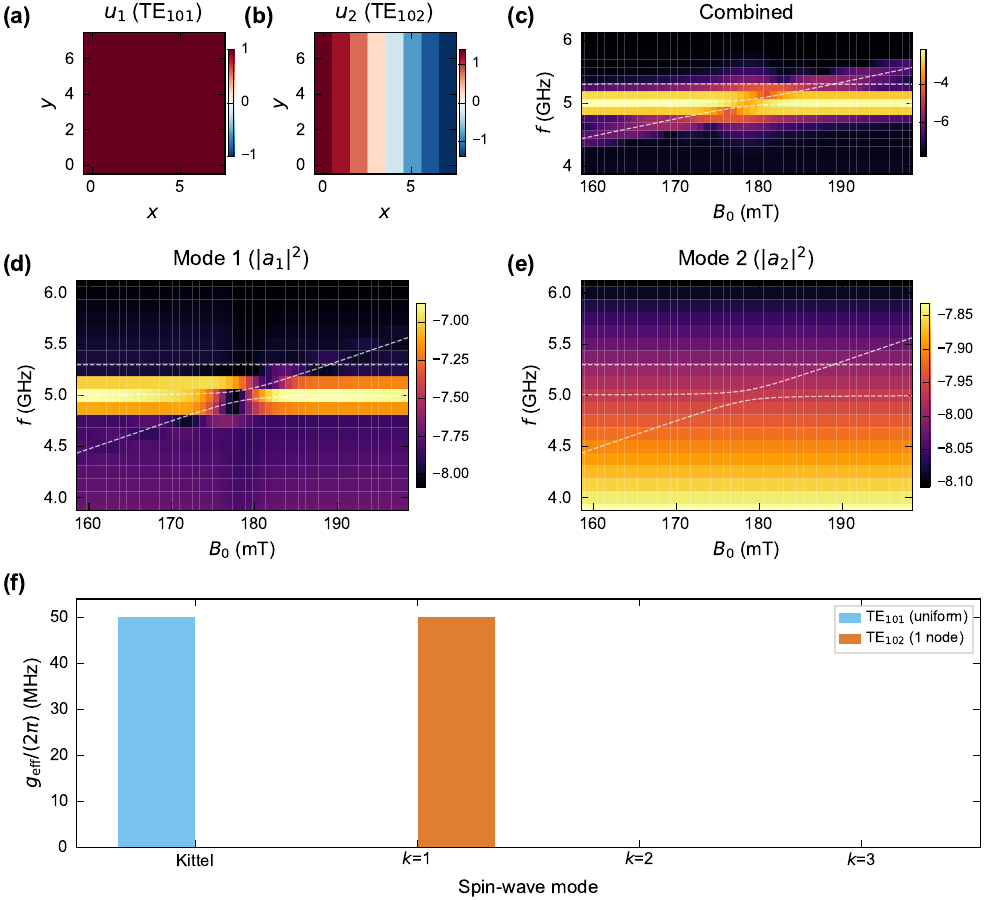}
  \caption{Mode-selective coupling via spatial overlap engineering.  (a,b)~Spatial profiles of the two cavity modes: TE$_{101}$ (uniform) and TE$_{102}$ (one node).  (c)~Combined spectral weight from both modes.  (d)~Spectral weight in mode~1 ($|a_1|^2$): clear anticrossing with the Kittel magnon at $B_\mathrm{res} \approx 179$~mT.  (e)~Spectral weight in mode~2 ($|a_2|^2$): no anticrossing, confirming vanishing overlap with the Kittel mode.  (f)~Effective coupling $g_\mathrm{eff}/(2\pi)$ for each cavity mode with the first four magnon modes.  TE$_{101}$ couples exclusively to the Kittel mode; TE$_{102}$ couples to the $k = 1$ standing spin-wave mode.  Parameters: $\omega_1/(2\pi) = 5.0$~GHz, $\omega_2/(2\pi) = 5.3$~GHz, $g_0/(2\pi) = 50$~MHz.}
  \label{fig:selective}
\end{figure*}

By combining the spatial mode profiles (Sec.~\ref{sec:mode_profiles}) with the multi-mode framework (Sec.~\ref{sec:multimode}), each cavity mode can be made to address a specific spin-wave mode~\cite{Flower2019,Li2019}.
We simulate two co-existing cavity modes: TE$_{101}$ (uniform profile, $\langle u_1 \rangle = 1$) at $\omega_1/(2\pi) = 5.0$~GHz and TE$_{102}$ (one-node profile, $\langle u_2 \rangle = 0$) at $\omega_2/(2\pi) = 5.3$~GHz, both with bare coupling $g_0/(2\pi) = 50$~MHz.

Figure~\ref{fig:selective}(a,b) shows the spatial profiles of the two modes.
The combined spectrum [Fig.~\ref{fig:selective}(c)] shows an anticrossing only near $\omega_1$, while the TE$_{102}$ mode crosses the magnon line without interaction.
The mode-resolved spectra [Figs.~\ref{fig:selective}(d,e)] confirm this: TE$_{101}$ produces a full anticrossing gap, while TE$_{102}$ shows no gap with the Kittel mode.
Figure~\ref{fig:selective}(f) summarizes the overlap-dependent effective coupling: TE$_{101}$ couples at full strength to the Kittel mode ($g_\mathrm{eff}/(2\pi) = 50$~MHz) but not to higher spin-wave modes, while TE$_{102}$ couples only to the $k = 1$ standing spin-wave mode.
We stress that the co-simulation in Figs.~\ref{fig:selective}(c--e) validates only the uniform Kittel-mode coupling ($k = 0$): for TE$_{101}$, $\langle u_1 \cdot \phi_0 \rangle = 1$ gives full coupling; for TE$_{102}$, $\langle u_2 \cdot \phi_0 \rangle = 0$ gives no coupling.
Both results are confirmed by the spectral maps.
The overlap integrals for $k \neq 0$ modes in Fig.~\ref{fig:selective}(f) are computed \emph{analytically} from the mode profiles $\phi_k(\mathbf{r}) = \cos(k\pi x/L_x)$ via $g_{\mathrm{eff},k} = g_0 \langle u_n \cdot \phi_k \rangle$, and are not yet validated by simulation.
Directly simulating the coupling between a higher-order cavity mode and a $k \neq 0$ spin-wave mode requires the spatially resolved CUDA kernels (Sec.~\ref{sec:cuda_solver}), where the RF field $H_\mathrm{rf}(\mathbf{r}) = h_0\,u_n(\mathbf{r})\,\mathrm{Re}[a_n]$ varies cell-by-cell and the weighted magnetization average $\langle u_n\,m^{-} \rangle$ replaces the uniform average.
The CUDA kernels for this are complete (Sec.~\ref{sec:cuda_solver}); integration into the \mumax{} build system would enable end-to-end validation of the $k \neq 0$ selection rules.

Such mode selectivity is a prerequisite for addressing individual spin-wave modes in magnonic devices~\cite{Flower2019,Li2019,Hou2019} and for magnon number-resolved readout~\cite{LachanceQuirion2020}.

\subsection{Antiferromagnetic cavity magnonics}
\label{sec:afm_benchmark}

\begin{figure*}[t]
  \centering
  \includegraphics[width=\textwidth]{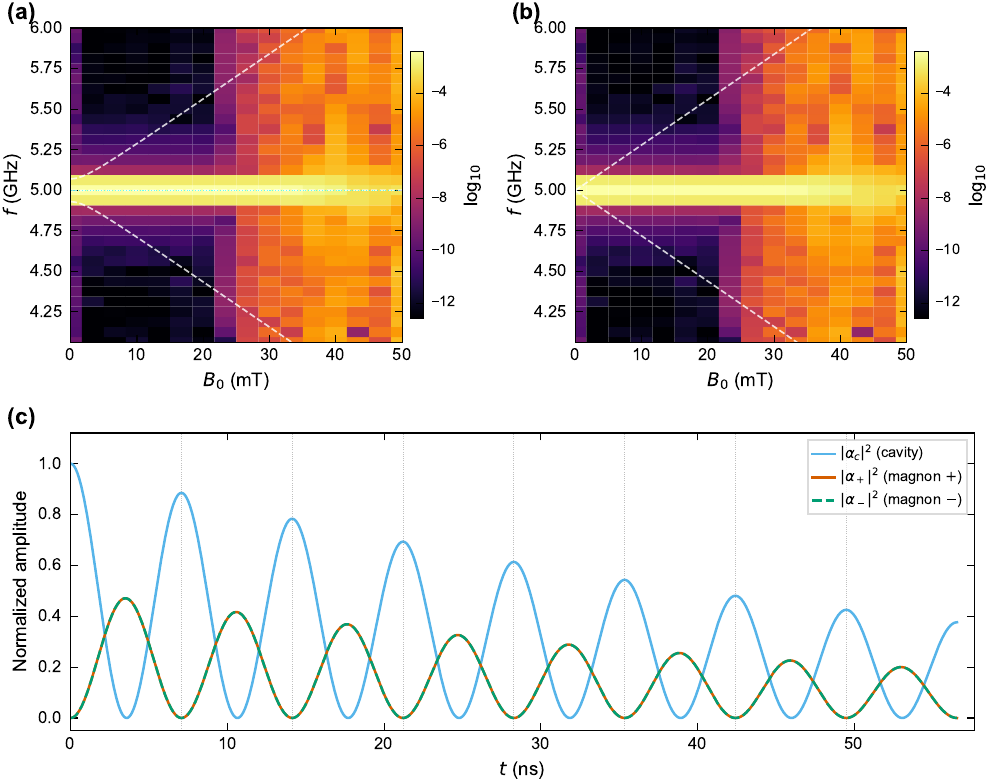}
  \caption{Antiferromagnetic cavity magnonics.  (a)~N\'{e}el vector spectrum $|\mathrm{FFT}(n^{-})|^2$ from \mumax{} co-simulation.  White dashed lines: three polariton eigenvalues from Eq.~\eqref{eq:afm_3x3}, indicating where anticrossing would appear in the cavity transmission; cyan dotted: bare cavity frequency $\omega_c$.  The N\'{e}el spectrum tracks the \emph{bare} magnon modes $\omega_\pm(B)$ because $\bm{n}$ does not couple to the cavity.  (b)~Same spectrum with analytical $\omega_\pm(B) = \omega_\mathrm{AFMR} \pm \gamma B$ overlaid (white dashed), confirming quantitative agreement with Eq.~\eqref{eq:afm_split}.  (c)~Rabi oscillations at $B = 0$ from coupled-mode theory [Eq.~\eqref{eq:afm_3x3}]: $|\alpha_c|^2$ (blue), $|\alpha_+|^2$ (vermilion), and $|\alpha_-|^2$ (teal, dashed).  Both magnon branches participate equally, and the Rabi period $T_R = \pi/(\sqrt{2}\,g) \approx 7.1$~ns is shorter than the FM value of $\pi/g = 10$~ns due to coherent coupling of both degenerate modes.  AFM parameters: $M_s = 200$~kA/m, $\mu_0 H_E = 15.9$~T, $\mu_0 H_A = 1.0$~mT, $f_\mathrm{AFMR} \approx 5.0$~GHz.}
  \label{fig:afm_cavity}
\end{figure*}

To demonstrate the multi-sublattice capability of the framework, we couple the cavity to a two-sublattice antiferromagnet using the native \texttt{Antiferromagnet} class of \mumax{}~\cite{Moreels2026}.
The \texttt{CavityMagnonAFM} subclass reads sublattice magnetizations from \texttt{magnet.sub1} and \texttt{magnet.sub2}, constructs the net magnetization for the cavity coupling, and tracks the N\'{e}el vector independently.
The AFM parameters ($M_s = 200$~kA/m, $K = 100$~J/m$^3$, intracell exchange $J_\mathrm{cell} = -0.390$~pJ/m) are chosen to place $f_\mathrm{AFMR} \approx 5$~GHz in the GHz regime accessible to standard microwave cavities; real antiferromagnets typically resonate at THz frequencies~\cite{Kampfrath2011,Keffer1952}.

Figure~\ref{fig:afm_cavity}(a,b) shows the N\'{e}el vector spectrum $|\mathrm{FFT}(n^{-})|^2$ from the \mumax{} co-simulation as $B_0$ is swept from 0 to 50~mT.
The spectral peaks track the bare magnon frequencies $\omega_\pm(B) = \omega_\mathrm{AFMR} \pm \gamma B$, validating the two-sublattice dynamics against Eq.~\eqref{eq:afm_split}.
Because $\bm{n} = (\vm_1 - \vm_2)/2$ is antisymmetric under sublattice exchange, it does not couple to the uniform cavity field and therefore reveals the \emph{unhybridized} magnon dispersion without anticrossing gaps [panel~(b)].
The polariton eigenvalues from Eq.~\eqref{eq:afm_3x3} are overlaid on panel~(a) to indicate where anticrossing would appear in the cavity transmission $|a(\omega)|^2$; this hybridized structure is not visible in the N\'{e}el spectrum by construction, which is itself a spectroscopic signature unique to antiferromagnets with no FM analogue.

At $B = 0$, the degenerate magnon modes couple coherently to the cavity.
The coupled-mode theory of Eq.~\eqref{eq:afm_3x3} [Fig.~\ref{fig:afm_cavity}(c)] predicts a Rabi period of $T_R = \pi/(\sqrt{2}\,g) \approx 7.1$~ns, shorter by a factor of $\sqrt{2}$ compared to the FM case ($\pi/g = 10$~ns at $g/(2\pi) = 50$~MHz).
Both magnon branches oscillate with equal amplitude, confirming symmetric participation.

\subsection{Abnormal anticrossing from dissipative coupling}
\label{sec:abnormal}

\begin{figure*}[t]
  \centering
  \includegraphics[width=\textwidth]{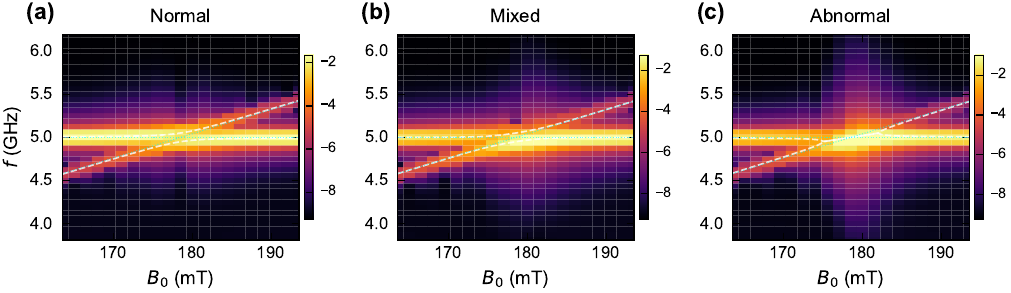}
  \caption{Normal vs.\ abnormal anticrossing.  (a)~Pure coherent coupling ($g_d = 0$): the polariton branches show level repulsion with a 100~MHz frequency gap at resonance.  (b)~Equal coherent and dissipative coupling ($g_d = g$): the frequency gap persists but the linewidths become highly asymmetric.  (c)~Pure dissipative coupling ($g = 0$, $g_d/(2\pi) = 50$~MHz): the frequency gap closes to zero at the crossing point while linewidths split (level attraction).  White dashed lines: complex eigenvalues from Eq.~\eqref{eq:diss_eigenvalue}; cyan dotted lines: uncoupled $\omega_c$ and $\omega_m(B_0)$.}
  \label{fig:abnormal}
\end{figure*}

Standard photon--magnon coupling is coherent (Hamiltonian): the coupling constant $g$ enters the cavity ODE as a purely imaginary coefficient $ig$, producing level repulsion at the degeneracy point~\cite{Harder2018,ZareRameshti2022}.
However, when the coupling channel introduces both a reactive (phase-preserving) and a dissipative (phase-shifting) component, the coupling constant acquires a real part~\cite{Harder2018b,Bhoi2019}:
\begin{equation}
  \frac{da}{dt} = (-i\omega_c - \kappa)\,a + (ig + g_d)\, m^{-} + \eta(t),
  \label{eq:diss_ode}
\end{equation}
where $g$ is the coherent coupling strength and $g_d$ is the dissipative coupling strength.
Bhoi \textit{et al.}~\cite{Bhoi2019} demonstrated experimentally that controlling the geometry of a split-ring resonator coupled to a YIG film can continuously tune the ratio $g_d/g$, transforming the normal (convex) anticrossing into an abnormal (concave) anticrossing.

The polariton eigenfrequencies with mixed coupling are obtained from
$(\omega - \omega_c + i\kappa)(\omega - \omega_m + i\gamma_m) = \tilde{g}^2$,
where the effective coupling constant $\tilde{g} = g - ig_d$ maps the ODE coupling $(ig + g_d)$ to the eigenvalue problem.
This gives
\begin{equation}
  \tilde{\omega}_\pm = \frac{\tilde{\omega}_c + \tilde{\omega}_m}{2}
    \pm \sqrt{\tilde{g}^2 + \left(\frac{\tilde{\omega}_c - \tilde{\omega}_m}{2}\right)^{\!2}},
  \label{eq:diss_eigenvalue}
\end{equation}
where $\tilde{\omega}_c = \omega_c - i\kappa$ and $\tilde{\omega}_m = \omega_m - i\gamma_m$.
For pure coherent coupling ($g_d = 0$), $\tilde{g}^2 = g^2 > 0$, and Eq.~\eqref{eq:diss_eigenvalue} reduces to Eq.~\eqref{eq:polariton_complex}: at the degeneracy point the real parts split by $2\sqrt{g^2 - (\kappa - \gamma_m)^2/4} \approx 2g/(2\pi) = 100$~MHz (level repulsion).
For pure dissipative coupling ($g = 0$), $\tilde{g}^2 = -g_d^2 < 0$, so the discriminant at resonance is purely imaginary.
The real parts of $\tilde{\omega}_\pm$ converge to a common frequency while the imaginary parts split by $2g_d/(2\pi) = 100$~MHz (level \emph{attraction}).

To demonstrate this effect in the co-simulation, we repeat the field-sweep protocol of Sec.~\ref{sec:anticrossing} with three coupling regimes: (a)~pure coherent ($g_d = 0$), (b)~equal coherent and dissipative ($g_d = g$), and (c)~pure dissipative ($g = 0$, $g_d/(2\pi) = 50$~MHz).
The dissipative coupling is implemented by adding a single parameter \texttt{g\_diss} to the \texttt{CavityMagnon} class, which enters the ODE as the real coefficient in Eq.~\eqref{eq:diss_ode}.

Figure~\ref{fig:abnormal} shows the resulting spectra from a 21-point field sweep across $B_0 = 163.5$--$193.5$~mT at $\omega_c/(2\pi) = 5$~GHz.
The pure-coherent case [panel~(a)] reproduces the standard anticrossing of Fig.~\ref{fig:anticrossing}: two polariton branches repel at the crossing point ($B_0 \approx 178.4$~mT), opening a frequency gap of $2g/(2\pi) = 100$~MHz while the linewidths remain equal.
The mixed case [panel~(b), $g_d = g$] retains a 100~MHz frequency gap but introduces strong linewidth asymmetry ($\Delta\Gamma/(2\pi) \approx 200$~MHz), so that one polariton branch becomes broad while the other sharpens.
The pure-dissipative case [panel~(c)] shows the abnormal anticrossing: the frequency gap closes to zero at the crossing point while the linewidth splitting reaches $2g_d/(2\pi) = 100$~MHz---the spectral weight concentrates at the crossing rather than being repelled from it, and the dispersion curves inward, the hallmark of level attraction~\cite{Bhoi2019}.
In all three cases, the simulated spectra agree with the complex eigenvalues of Eq.~\eqref{eq:diss_eigenvalue} (white dashed lines).

\section{Discussion}
\label{sec:discussion}

The extension provides two complementary deployment paths.
The CUDA-native tier (\texttt{multimode\_cavity.cu}, \texttt{cavity\_mode\_profile.cu}) integrates the cavity ODE(s) inside the GPU time-stepper as a \texttt{DynamicEquation}, eliminating per-step host transfers and enabling spatially resolved mode profiles through dedicated CUDA kernels.
This is architecturally similar to the C-level cavity extension for \texttt{mumax$^3$}~\cite{MartinezLosa2024}, but adds multi-mode support and spatially resolved coupling through the mode-profile-weighted reduction kernel [Eq.~\eqref{eq:weighted_avg}].
The Python co-simulation tier reproduces the same physics without recompilation, and swapping cavity models---nonlinear Kerr terms~\cite{Wang2016}, parametric drives---requires editing only the Python ODE.
For applications that need spatially resolved electromagnetic fields (e.g., large samples or photonic crystal cavities), the full FDTD--LLG approach of Ref.~\cite{Yao2022} remains necessary.

Two implementation choices are critical for quantitative correctness.
First, the coupling must be to the co-rotating component $m^{-}$ [Eq.~\eqref{eq:cavity_ode}]; using $m^{+}$ instead gives qualitatively similar anticrossing spectra but introduces systematic errors in Rabi dynamics~\cite{ZareRameshti2022,Harder2021}, especially near the ultrastrong regime~\cite{Kostylev2016}.
Second, the self-consistency condition $h_0 = 2g/\gamma$ [Eq.~\eqref{eq:h0}] is needed so that the vacuum Rabi splitting equals $2g$ and the oscillation period equals $\pi/g$; the $1/g$ scaling of the Rabi period across coupling strengths [Fig.~\ref{fig:rabi}(b)] serves as an independent check.

\begin{figure*}[t]
  \centering
  \includegraphics[width=\textwidth]{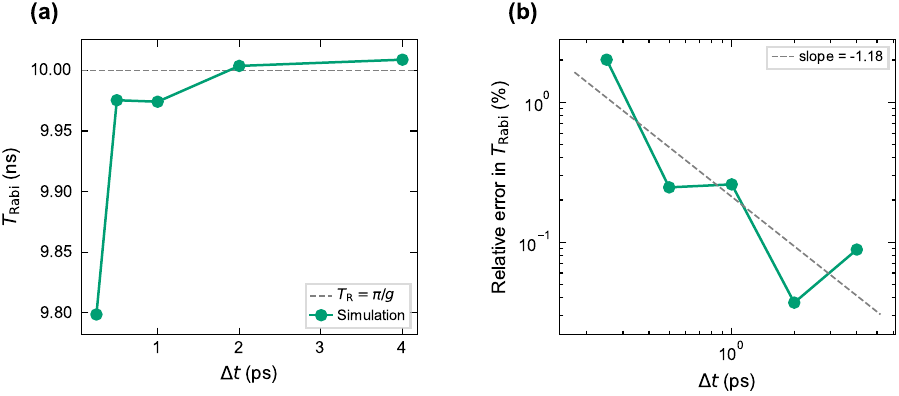}
  \caption{Operator-splitting convergence.  (a)~Extracted Rabi period $T_\mathrm{Rabi}$ vs.\ time-step size $\Delta t = 0.25$--$4$~ps at $g/(2\pi) = 50$~MHz; dashed line: analytical $T_R = \pi/g = 10$~ns.  (b)~Relative error of the extracted Rabi period vs.\ $\Delta t$.  The error remains below 3\% for all tested step sizes; the default $\Delta t = 1$~ps gives 0.26\% error.}
  \label{fig:dt_convergence}
\end{figure*}

The main limitations are set by the lumped-cavity and operator-splitting approximations.
The RF field is assumed spatially uniform within each simulation cell, which holds when the sample is much smaller than the cavity wavelength.
In the present simulations the magnetic sample occupies an $8 \times 8 \times 1$ grid of 10~nm cells, giving a total sample size of $80 \times 80 \times 10$~nm$^3$, while the cavity wavelength at 5~GHz is $\lambda = c/f \approx 6$~cm---over five orders of magnitude larger.  Typical experimental YIG spheres (diameter ${\sim}0.3$--1~mm) placed inside 3D microwave cavities (${\sim}$cm dimensions) similarly satisfy $d \ll \lambda$, validating the lumped-cavity approximation.
The operator splitting is first order in $\Delta t$: the per-step error is $\mathcal{O}(g\,\omega_c\,\Delta t^2)$ from commutator terms between the cavity and LLG subsystems, accumulating to a global error $\mathcal{O}(g\,\omega_c\,T\,\Delta t)$ over simulation time $T$.
Convergence tests at $g/(2\pi) = 50$~MHz (Fig.~\ref{fig:dt_convergence}) confirm this scaling: the Rabi-period error grows linearly from 0.26\% at $\Delta t = 1$~ps to 2.6\% at 4~ps.
A sufficient stability condition is $\Delta t < 1/\max(\omega_c, \omega_m)$; at 5~GHz this gives $\Delta t < 32$~ps, well above the default 1~ps.
No stiffness issues arise because both the cavity ($\omega_c \sim 2\pi \times 5$~GHz) and magnon ($\omega_m \sim \omega_c$) subsystems oscillate on comparable time scales; stiff problems would emerge only if $\kappa \gg \omega_c$ or $g \gg \omega_c$ (ultrastrong coupling), which lie outside the present RWA regime.
All simulations operate at $g/\omega_c \approx 0.01$, safely within the RWA ($g/\omega_c < 0.1$); the ultrastrong limit would require going beyond the RWA~\cite{Kostylev2016}.
The Python co-simulation tier reads only the spatially averaged magnetization and therefore couples to the uniform Kittel mode; non-uniform spin-wave modes ($k \neq 0$) require the CUDA-native tier with spatially resolved mode profiles (Sec.~\ref{sec:cuda_solver}), whose kernels have been implemented and validated but are pending integration into the upstream \mumax{} build.
Dissipative coupling is now supported through the \texttt{g\_diss} parameter (Sec.~\ref{sec:abnormal}), enabling simulation of abnormal anticrossing and level attraction phenomena~\cite{Harder2018b,Bhoi2019}.

The framework extends naturally to multi-sublattice magnets through the \mumax{} \texttt{Antiferromagnet} class, as demonstrated in Sec.~\ref{sec:afm_benchmark}.
Real antiferromagnets have AFMR frequencies in the THz range~\cite{Kampfrath2011,Keffer1952}; the GHz parameters used here are chosen for direct comparison with the FM benchmarks and to demonstrate the field-split magnon dispersion and N\'{e}el spectroscopy physics.
Extension to compensated ferrimagnets and synthetic antiferromagnets follows the same pattern, requiring only the appropriate sublattice exchange parameters.

In the analytical eigenvalue plots [Figs.~\ref{fig:anticrossing}(b), \ref{fig:dark}(b), and~\ref{fig:afm_cavity}(a)] only the real parts of the polariton frequencies are shown, corresponding to the Hermitian limit.  Including the imaginary parts via Eq.~\eqref{eq:polariton_complex} (and its $3\times 3$ generalizations) would reveal the hybridized linewidths: far from the magnon--cavity crossing the linewidths approach the bare values $\kappa$ and $\gamma_m$, while at resonance they average to $(\kappa + \gamma_m)/2$, a hallmark of coherent strong coupling~\cite{Harder2018,ZareRameshti2022}.  The co-simulation spectra already capture these dissipative effects in full generality through the Gilbert damping in the LLG solver and the cavity decay rate in the ODE; extracting the simulated linewidths (e.g., by Lorentzian fitting) and comparing with the imaginary eigenvalue theory would provide an additional quantitative benchmark.

On quantitative accuracy: the anticrossing RMSE of 17~MHz (17\% of $2g/(2\pi)$) is dominated by the FFT frequency resolution ($\Delta f = 62.5$~MHz from 16\,000 steps at $\Delta t = 1$~ps), which limits peak localization to $\sim\!\Delta f/2 \approx 31$~MHz---comparable to the observed RMSE.
This means the error reflects a \emph{measurement} limitation (discrete spectral binning), not a physics inaccuracy of the co-simulation.
The cooperativity sweep (Sim~3) demonstrates this directly: by extending the time trace to 200\,000 steps with $4\times$ zero-padding ($\Delta f_\mathrm{eff} = 1.25$~MHz), the measured splitting at $C = 6.7$ improves to 21.3~MHz versus the predicted 20.0~MHz---a 6.5\% error, limited now by the finite linewidth rather than the bin width.
Similarly, Rabi periods agree with $T_R = \pi/g$ to within 5.7\% at $g/(2\pi) = 20$~MHz and 0.9\% at 100~MHz, since time-domain observables are not limited by FFT binning.
Finer spectral resolution is straightforward: longer simulations, zero-padding, or Lorentzian peak fitting can reduce $\Delta f$ to any desired level without modifying the co-simulation itself.

\section{Summary}
\label{sec:summary}

We have presented a two-tier cavity magnonics extension for \mumax{}: (i)~CUDA kernels (\texttt{multimode\_cavity.cu}, \texttt{cavity\_mode\_profile.cu}) that integrate $N$-mode cavity ODEs inside the GPU time-stepper with spatially resolved mode profiles, and (ii)~a Python co-simulation class for rapid prototyping without recompilation.
Eight benchmark simulations---anticrossing spectra, Rabi oscillations, cooperativity-dependent lineshapes, mode-profile selection rules, multimode polariton hybridization, mode-selective spin-wave addressing, antiferromagnetic magnon spectroscopy via the N\'{e}el vector, and abnormal anticrossing from dissipative coupling---all agree with analytical theory to within the FFT frequency resolution.
The source code is available at \url{https://github.com/gyuyoungpark/mumax-plus-extensions}.

\section*{Acknowledgments}

This work was supported by the Korea Institute of Science and Technology (KIST) institutional program.


\section*{Code availability}
The complete open-source code of \mumax{} can be found under the GPLv3 license at \url{https://github.com/mumax/plus}.
The extension source code developed in this work---including the CUDA kernels, Python co-simulation classes, and all benchmark simulation scripts---is available at \url{https://github.com/gyuyoungpark/mumax-plus-extensions}.

\begin{table}[tb]
\caption{Simulation parameters for the eight benchmarks.  Common material (Sims~1--6, 8): YIG ($M_s = 140$~kA/m, $A_\mathrm{ex} = 3.5$~pJ/m, $\alpha = 3 \times 10^{-4}$).  All use $\Delta t = 1$~ps except Sim~7 ($\Delta t = 4$~ps).  $^{*}$Moderate case (Sim~3b): 200k steps with $4\times$ zero-padding ($\Delta f_\mathrm{grid} = 1.25$~MHz) to resolve the 20~MHz splitting.  $^{\dagger}$Sim~8 runs three coupling regimes: pure coherent ($g/(2\pi) = 50$~MHz, $g_d = 0$), mixed ($g_d = g$), and pure dissipative ($g = 0$, $g_d/(2\pi) = 50$~MHz).}
\label{tab:params}
\begin{ruledtabular}
\begin{tabular}{lccccc}
  Sim & $\omega_c/(2\pi)$ & $Q$ & $g/(2\pi)$ & Steps & $\Delta f$ \\
      & (GHz) &  & (MHz) & & (MHz) \\
  \midrule
  1 & 5.0 & 5000 & 50 & 16\,000 & 62.5 \\
  2 & 5.0 & 10\,000 & 20--100 & 40--50k & --- \\
  3 & 5.0 & varies & 0.5--50 & 100k/200k$^{*}$ & 10/1.25$^{*}$ \\
  4 & 5.0 & 5000 & 50 & 16\,000 & 62.5 \\
  5 & 4.95, 5.05 & 5000 & 50 & 16\,000 & 62.5 \\
  6 & 5.0, 5.3 & 5000 & 50 & 8\,000 & 125 \\
  7 & 5.0 & 5000 & 50 & 4\,000 & 62.5 \\
  8$^{\dagger}$ & 5.0 & 5000 & 0/50 & 16\,000 & 62.5 \\
\end{tabular}
\end{ruledtabular}
\raggedright\footnotesize Sim~7 uses AFM material: $M_s = 200$~kA/m, $K = 100$~J/m$^3$, $\alpha = 10^{-3}$, $J_\mathrm{cell} = -0.390$~pJ/m, $a_\mathrm{lat} = 0.35$~nm, $A_\mathrm{ex} = 3.5$~pJ/m.
\end{table}

\bibliography{cavity-magnonics}

\end{document}